\documentclass{article}
\usepackage{spconf,amsmath,graphicx,physics,cite,amssymb,xcolor,subcaption,booktabs,bm,dblfloatfix}

\title{JOINT OPTIMIZATION OF STREAMING AND NON-STREAMING AUTOMATIC SPEECH RECOGNITION WITH MULTI-DECODER AND KNOWLEDGE DISTILLATION}
\name{Muhammad Shakeel$^1$, Yui Sudo$^1$, Yifan Peng$^2$, Shinji Watanabe$^2$}
\address{
  $^1$Honda Research Institute Japan Co., Ltd., Saitama, Japan\\
  $^2$Carnegie Mellon University, Pittsburgh, PA, USA}
\begin{document}
\ninept
\maketitle
\begin{abstract}
End-to-end (E2E) automatic speech recognition (ASR) can operate in two modes: streaming and non-streaming, each with its pros and cons. Streaming ASR processes the speech frames in real-time as it is being received, while non-streaming ASR waits for the entire speech utterance; thus, professionals may have to operate in either mode to satisfy their application. In this work, we present joint optimization of streaming and non-streaming ASR based on multi-decoder and knowledge distillation. Primarily, we study 1) the encoder integration of these ASR modules, followed by 2) separate decoders to make the switching mode flexible, and enhancing performance by 3) incorporating similarity-preserving knowledge distillation between the two modular encoders and decoders. Evaluation results show 2.6\%-5.3\% relative character error rate reductions (CERR) on CSJ for streaming ASR, and 8.3\%-9.7\% relative CERRs for non-streaming ASR within a single model compared to multiple standalone modules.
\end{abstract}
\section{Introduction}
The focus of modern end-to-end (E2E) automatic speech recognition (ASR) systems is to explore exemplary architectures \cite{2023endtoend} and surpass conventional ASR systems. ASR encompasses various tasks, including streaming \cite{streaming1,streaming2} and high-performance non-streaming \cite{conformer1}, demonstrating minimum latency \cite{streaming3,streaming4} or word error rate \cite{conformer2} in each task. However, creating specialized architectures for each possible task \cite{ctc1,ctc2,rnnt1,attention1,attention2,nar1}, leads to an enormous catalog of models making them less scalable. Eventually, it is desirable for E2E-ASR systems to be highly accurate, have low latency, and are unifiable which makes their deployment easy for various real-world speech applications. To get there, we explore the potential of multi-task learning \cite{multitask1} and knowledge distillation \cite{kdonline2} and adopt a single E2E-ASR model development route. We ask the following question: \textit{can inter-module knowledge exchange within a single model fundamentally elevate the performance capabilities of the ASR system?} In simple words, can we use the hidden states of the streaming encoder as an input to the non-streaming encoder while simultaneously using the hidden states as an auxiliary knowledge distillation loss. As a general approach rather than only relying on the cascaded integration with shared RNN-T decoder \cite{cascaded0} for performance optimization, we propose to use multiple decoders (CTC \cite{ctc1}, RNN-T \cite{rnnt1}, attention \cite{attention1},  and masked language model (MLM) \cite{nar1}) within a single model and additionally employ similarity-preserving knowledge distillation (sp-KD) technique \cite{kd1} for both encoder and decoder modules, which guides the training of the streaming (student) network such that the hidden encoder and decoder states that produce similar (dissimilar) representations in the non-streaming (teacher) network produce similar (dissimilar) representations in the student network.
\begin{figure}[t!]
\begin{center}
{\resizebox*{8.5cm}{!}{\includegraphics[width=\textwidth]{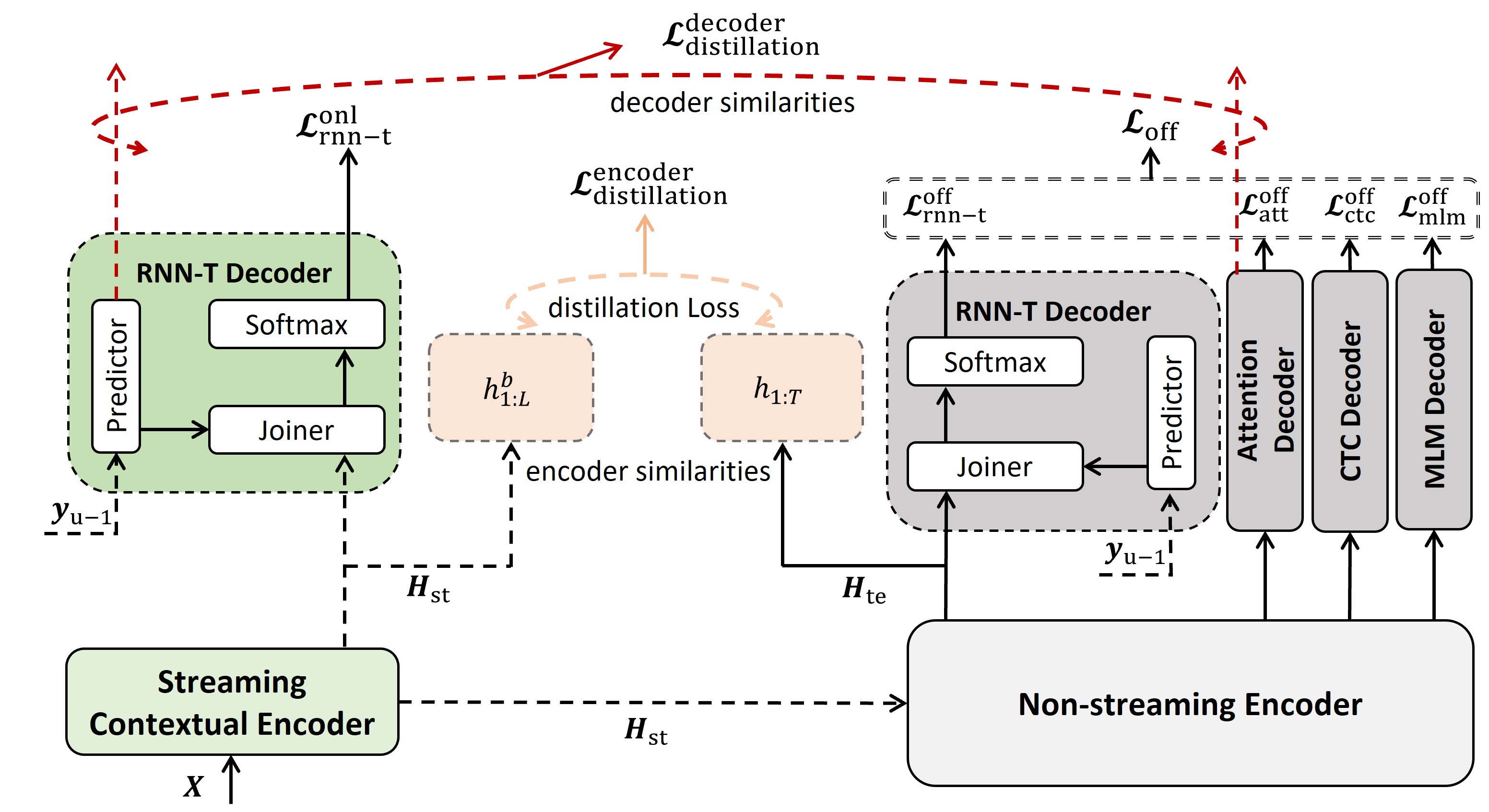}}}
\vspace*{-2mm}
\caption{Joint optimization of multi-decoder ASR model: A single model with streaming (student) and non-streaming (teacher) modules, both of which are jointly optimized.}
\label{cascaded} 
\end{center}
\vspace*{-10mm}
\end{figure}
Recently, there has been a concentrated effort to unify streaming and non-streaming ASR models \cite{joint2,cascaded4,parallel}. The streaming ASR processes information within a limited context, whereas the non-streaming ASR leverages full-context modeling resulting in unique acoustic features for each of them. Given the distinctive contextual information inherent to each network architecture, it is not straight forward to perform joint optimization within a single network architecture; thus, resulting in separate models for each application scenario. Former methodologies \cite{joint2,cascaded4,parallel,parallel2,kd3,kd4,joint1,cascaded1,cascaded3,kd1,multitask3} seeking to unify streaming and non-streaming ASR predominantly concentrated on single decoders, such as RNN-T, and employed distillation techniques using KL-divergence or mean square error (MSE). The performance of these methods when presented with multi-decoder architectures and sp-KD remains ambiguous. The broader field of simplifying and unifying ASR architectures has a variety of approaches for integrating limited and full context. Earlier work to unify streaming and non-streaming models used joint training and knowledge distillation, as our approach does, but later works with cascaded architecture \cite{cascaded1} showed better performance with shared RNN-T decoders. More recent work with cascaded architecture \cite{cascaded3} for sub-model training used separate RNN-T-based decoders to reduce the model size and improve accuracy. Our work also sits in the broader field of unifying architectures, introducing the idea that hidden representations of streaming and non-streaming encoders can be used as a sp-KD loss \cite{kd1}. Moreover, as opposed to using shared RNN-T decoder \cite{cascaded1}, we propose the use of separate decoders i.e., single RNN-T decoder for streaming path and four-decoder-based joint model \cite{multitask3} (i.e., using CTC, RNN-T, attention and Mask-CTC) for non-streaming path to effectively improve the performance of both modules and bring more regularization, improved robustness and flexibility in the E2E-ASR system.
\vspace*{-2mm}
\section{Joint optimization of ASR model}
This section introduces the proposed jointly optimized ASR model, followed by detailed descriptions of each of our design modules.

\textbf{Streaming and non-streaming encoder.}
In the proposed joint architecture, we use an effective block processing-based streaming encoder \cite{streaming-encoder1,streaming-encoder2} in a stack of $M$ encoder layers, connected by a non-streaming conformer encoder \cite{conformer1} with $N$ layers followed by separate decoders for each encoder path. We allow enhanced feature learning from hidden encoded features and embedded multiple decoders, as shown in Figure \ref{cascaded}. For streaming encoder, we use context inheritance mechanism introduced in \cite{streaming-encoder1}. Here, authors tend to retain the previous context using context embedding computed for each block at each sublayer and then forwarded it to the next sublayer. From Eq. \eqref{eq:block}, we extract the sub-sampled hidden encoder states computed from the context embeddings using the contextual blocks. Let \begin{math}L_{\text{block}}\end{math} and \begin{math}L_{\text{hop}}\end{math} represent the block size and hop length, respectively. The \textit{b}-th block of the input audio feature sequence $\bm{X}^b$ is defined as, 
\begin{equation}
\bm{X}^{b} = (\bm{X}_t | t= (b-1)L_{\text{hop}} + 1, ..., (b-1)L_{\text{hop}} + L_{\text{block}} + 1)
\label{eq:block}
\end{equation}
The hidden state for each block, labeled as the $b$-th block, is encoded whereas each block contains a series of hidden states of \begin{math}L_{\text{block}}\end{math}- \textit{length}, i.e., \begin{math}\bm{H}^{b} = (\bm{h}_{1}^{b},\ldots,\bm{h}_{L_{block}}^{b}) \end{math}. This encoding procedure is applied in a sequential manner, ultimately resulting in hidden states of length $T$. We feed these features extracted from Eq. \eqref{eq:contextencoder} as an input to the non-streaming encoder where this feature sequence is transformed to a sub-sampled $T$-length hidden state sequence as shown in Eq. \eqref{eq:conformerencoder}.
\begin{equation}
\bm{H}_{\text{st}} = \mathrm{STConEncoder}(\bm{X}).
\label{eq:contextencoder}
\end{equation} 
\begin{equation}
\bm{H}_{\text{te}} = \mathrm{NSTConEncoder}(\bm{H}_{\text{st}}).
\label{eq:conformerencoder}
\end{equation}
In this work, the streaming contextual conformer encoder operates as a separate streaming encoder-decoder module. It is connected to a non-streaming encoder-decoder module through an output from the streaming encoder.

\textbf{Multi-decoder and knowledge distillation.}
Recently proposed cascaded encoders with a single shared decoder \cite{cascaded1} learns the hidden representations between the streaming and non-streaming context; however, using this approach leads to performance degradation, as studied in \cite{cascaded3} in either of the models. This is because the model is being optimized excessively for the streaming module and may need more capacity to fully capture the future contextual information required for the non-streaming module. As a result, the shared decoder architecture unintentionally applies a penalty on the non-streaming decoder and degrades its performance. One way to address this problem is to allow flexibility in the cascaded model architecture by keeping all the decoders separate. Inspired by \cite{cascaded3}, as opposed to the original cascaded structure \cite{cascaded1}, we leverage the separate decoders architecture to reduce the tension between the loss objectives. In addition, we also perform decoder-side distillation and bring more regularization for both the modules. Employing multiple decoders enhances the performance by leveraging the unique competencies of each decoder and facilitating comprehensive knowledge transfer. This approach increases the ASR model's adaptability to diverse applications and also improves its overall accuracy and robustness. Furthermore, the separate multi-decoder architecture allows us to leverage the increased weight assignment to the streaming decoder, keeping the performance at par for the non-streaming decoders and making the knowledge distillation from the offline to the online module more regularized. Since there are multiple input processing paths which makes the multi-task learning objective consists of multiple loss components as presented below:

\textbf{Streaming module loss.} The input acoustic features denoted by
\begin{math}\bm{X} = (x_{1},\ldots,x_{T})\end{math}, are first passed to the online streaming module that uses contextual block conformer \cite{streaming-encoder1} as an encoder and the RNN-T as a decoder. The streaming RNN-T decoder optimizes the model parameters by minimizing the negative log-likelihood given by:
\begin{equation}
\mathcal{L}_{\text{onl}} = -\sum_{( \bm{X} \rightarrow \bm{H_{\text{st}}}, \bm{y} )} \log P_{\text{onl}}(\bm{y} \mid \bm{H_{\text{st}}}).
\label{eq:rnnt_onl_loss}
\end{equation}

\textbf{Non-streaming module loss.} The offline non-streaming module is implemented using a shared full-context conformer block as an encoder with four different decoders i.e., CTC, RNN-T, attention mechanism and MLM to jointly optimize the offline multi-task loss. The offline architecture is adopted from \cite{multitask3} and comprises of following loss objectives as described in Eqs. \eqref{eq:ctc_off_loss}, \eqref{eq:rnnt_off_loss}, \eqref{eq:att_off_loss} and \eqref{eq:mlm_off_loss}:

The CTC decoder in the non-streaming module loss refines the model parameters by minimizing the negative log-likelihood given by:
\begin{equation}
L_{\text{ctc}} = -\sum_{( \bm{H_{\text{st}}} \rightarrow \bm{H_{\text{te}}}, \bm{y} )} \log P_{\text{ctc}}(\bm{y} \mid \bm{H_{\text{te}}}),
\label{eq:ctc_off_loss}
\end{equation}
While CTC adopts the conditional independence in Eq. \eqref{eq:ctc_off_loss}, the RNN-T decoder optimizes the model parameters by minimizing the negative log-likelihood given by:
\begin{equation}
\mathcal{L}_{\text{rnnt}} = -\sum_{( \bm{H_{\text{st}}} \rightarrow \bm{H_{\text{te}}}, \bm{y} )} \log P_{\text{rnnt}}(\bm{y} \mid \bm{H_{\text{te}}}),
\label{eq:rnnt_off_loss}
\end{equation}
Moreover, the attention decoder refines model parameters by minimizing the corresponding negative log-likelihood given by:
\begin{equation}
L_{\text{att}} = -\sum_{( \bm{H_{\text{st}}} \rightarrow \bm{H_{\text{te}}}, \bm{y} )} \log P_{\text{att}}(\bm{y} \mid \bm{H_{\text{te}}}).
\label{eq:att_off_loss}
\end{equation}
Finally, the MLM decoder estimates the token sequence using the full sequence given $\bm{H}_{\text{te}}$ in Eq. \eqref{eq:conformerencoder}, analogous to the attention case. However, during training, MLM differs from attention by masking randomly sampled tokens, \text{$y_{\text{mask}}$}, with a special token \text{$\text{$<$mask$>$}$}. Then, \text{$y_{\text{mask}}$} is predicted based on the remaining unmasked tokens, \text{$y_{\text{obs}}$}, as \begin{math} P_{\text{mlm}}(y_{\text{\text{mask}}} | y_{\text{obs}}, \bm{H_{\text{te}}}).\end{math} MLM refines model parameters by minimizing the negative log-likelihood given by:
\begin{equation}
L_{\text{mlm}} = -\sum_{( \bm{H_{\text{st}}} \rightarrow \bm{H_{\text{te}}}, \bm{y} )} \log P_{\text{mlm}}(y_{\text{mask}}|y_{\text{obs}}, \bm{H_{\text{te}}}).
\label{eq:mlm_off_loss}
\end{equation}
Finally, the offline loss \begin{math} (\mathcal{L}_{\text{off}})\end{math} is computed using the weighted sum of individual loss objectives from Eqs. \eqref{eq:ctc_off_loss}, \eqref{eq:rnnt_off_loss}, \eqref{eq:att_off_loss} and \eqref{eq:mlm_off_loss}:
\begin{equation}
\mathcal{L}_{\text{off}} = \lambda_{\text{ctc}} \mathcal{L}_{\text{ctc}} + \lambda_{\text{rnnt}} \mathcal{L}_{\text{rnnt}} + \lambda_{\text{att}} \mathcal{L}_{\text{att}} + \lambda_{\text{mlm}} \mathcal{L}_{\text{mlm}},
\label{eq:loss_offline}
\end{equation}
where \begin{math} \lambda_{\text{ctc}}, \lambda_{\text{rnnt}}, \lambda_{\text{att}}\end{math} and \begin{math} \lambda_{\text{mlm}} \end{math} are tunable hyperparameters and are determined experimentally. However, for this work we used the hyperparameters as reported in \cite{multitask3} and obtained optimal results.

\textbf{Similarity preserving knowledge distillation loss.}
We explore the idea of similarity-preserving knowledge distillation (sp-KD) initially proposed in computer vision \cite{kd1} to preserve the similarities between the hidden representations. We apply this concept of sp-KD for jointly optimizing the streaming and non-streaming ASR modules. In Figure \ref{cascaded}, we first look at the sp-KD loss that moves from the attention-based decoder to the RNN-T-based predictor, a process we call decoder-side distillation (sp-DD). Subsequently, we examine knowledge transfer from non-streaming hidden representations to the streaming hidden representations what is known as encoder-side distillation (sp-ED). We explain the concept of sp-KD in-terms of encoder layers, where we compute the similarities between the activations produced in both the encoder layers.  Given an input mini-batch, we denote the encoder activations produced by the offline (teacher) module for given layer \begin{math}l\end{math} as \begin{math} H_{\text{te}}^{(l)} \in R^{B\times T \times D} \end{math}, where \begin{math} B \end{math} is the batch size, \begin{math}T\end{math} is the length of the input audio sequence and \begin{math}D\end{math} is the dimensionality of each feature vector. For online (student) module, we denote the activations for the corresponding layer \begin{math}l^{'}\end{math} as \begin{math}H_{\text{st}}^{(l^{'})} \in R^{B^{'} \times T^{'} \times D^{'}} \end{math}. The hypothesis is to extract activation correlations from the non-streaming encoder and guide the streaming encoder toward these activations. We first take the L2-normalized outer products of the induced streaming (\begin{math} H_{\text{te}}^{(l)}\end{math}) and non-streaming (\begin{math}H_{\text{st}}^{(l^{'})} \end{math}) encoder activations and define a distillation loss that penalizes the differences between the learned representations of both the encoder output sequences given as:
\begin{equation}
\Tilde{G}_{\text{te}}^{(l)} = Q_{\text{te}}^{(l)} \cdot Q_{\text{te}}^{(l)^{\intercal}}; \quad G_{\text{te}[i,:]}^{(l)} = \Tilde{G}_{\text{te}[i,:]}^{(l)} / \norm{\Tilde{G}_{\text{te}[i,:]}^{(l)}}_{2}
  \label{eq:act_encoder_t}
\end{equation}
\begin{equation}
\Tilde{G}_{\text{st}}^{(l^{'})} = Q_{\text{st}}^{(l^{'})} \cdot Q_{\text{st}}^{(l^{'})^{\intercal}}; \quad G_{\text{st}[i,:]}^{(l^{'})} = \Tilde{G}_{\text{st}[i,:]}^{(l^{'})} / \norm{\Tilde{G}_{\text{st}[i,:]}^{(l^{'})}}_{2}
  \label{eq:act_encoder_s}
\end{equation}
where in equation \eqref{eq:act_encoder_t} and \eqref{eq:act_encoder_s}, we perform reshaping of \begin{math} H_{\text{te}}^{(l)} \end{math} and \begin{math} H_{\text{st}}^{(l^{'})}  \end{math}, i.e., \begin{math} Q_{\text{te}}^{(l)} \in R^{B\times TD} \end{math} and \begin{math} Q_{\text{st}}^{(l^{'})} \in R^{B^{'} \times T^{'}D^{'}} \end{math}, resulting in \begin{math} \Tilde{G}_{\text{te}}^{(l)} \end{math} and \begin{math} \Tilde{G}_{\text{st}}^{(l^{'})} \end{math} as a \begin{math} B \times B \end{math} matrix. Moreover, we apply row wise L2 normalization and obtain \begin{math} G_{\text{te}[i,:]}^{(l)} \end{math} and \begin{math} G_{\text{st}[i,:]}^{(l^{'})} \end{math} respectively, where the notation [\textit{i},:] denotes the \textit{i}th row in a matrix. Thus, the overall sp-KD loss can be defined as 
\begin{equation}
\mathcal{L}_{\text{dist}}(G_{\text{te}},G_{\text{st}}) = \frac{1}{B^{2}} \sum_{(l,l^{'}) \in \zeta } \norm{G_{\text{te}}^{(l)} - G_{\text{st}}^{(l^{'})}}_{F}^{2}
  \label{eq:loss_dist}
\end{equation}
where \begin{math} \zeta \end{math} adds the \begin{math} (l,l^{'}) \end{math} layer pair similarities at the end of each encoder block. In Eq. \eqref{eq:loss_dist} \begin{math}, \norm{\cdot}_{F}^{2} \end{math} is the Frobenius norm, and performs a mean element-wise squared difference between the \begin{math} G_{\text{te}}^{(l)} \end{math} and \begin{math} G_{\text{st}}^{(l^{'})} \end{math} matrices.

Finally, we define the total multi-task learning objective using knowledge distillation as: 
\begin{equation}
\mathcal{L}_{\text{mtl}} = \lambda_{\text{onl}}\mathcal{L}_{\text{onl}} + \lambda_{\text{off}}\mathcal{L}_{\text{off}} + \lambda_{\text{dist}}\mathcal{L}_{\text{dist}}
  \label{eq:mtl_loss}
\end{equation}
where \begin{math} \lambda_{\text{onl}}, \lambda_{\text{off}} \end{math} and \begin{math}\lambda_{\text{dist}}\end{math} are the weighting terms and \begin{math} \mathcal{L}_{\text{onl}}\end{math} is the online loss obtained from the streaming path in Eq. \eqref{eq:rnnt_onl_loss} and \begin{math} \mathcal{L}_{\text{off}}\end{math} is the offline loss obtained from the non-streaming path in Eq. \eqref{eq:loss_offline} and  \begin{math} \mathcal{L}_{\text{dist}}\end{math} is the knowledge distillation loss obtained from the intermediate representations of the streaming and non-streaming encoders or decoders in Eq. \eqref{eq:loss_dist}.
\vspace*{-3mm}
\begin{table}[b!]
\vspace*{-3mm}
\caption{On Corpus of Spontaneous Japanese (CSJ) \cite{csj}: Absolute (abs.) character error rate (CER) and relative (rel.) CERR numbers on CSJ data. Best CER result is \textbf{bolded} and best overall results are further \underline{\textbf{underlined}}.}
\vspace*{-5mm}
\label{ablation1}
\begin{center}
\scalebox{0.72}{
\begin{tabular}{@{}llc@{\hspace{0.1cm}}c@{\hspace{0.5cm}}c@{\hspace{0.1cm}}c@{\hspace{0.5cm}}c@{\hspace{0.1cm}}c}
\toprule
                    &                                 & \multicolumn{6}{c}{\textbf{CSJ (SUBSET A)}}  \\
\cmidrule{3-8}
\textbf{ID}         & \textbf{Method}                 & \multicolumn{2}{c}{\textbf{eval1}}              & \multicolumn{2}{c}{\textbf{eval2}}           &  \multicolumn{2}{c}{\textbf{eval3}}  \\
\cmidrule{3-8}
                    &                                 &  abs.$\downarrow$   & rel.(\%)$\uparrow$ & abs.$\downarrow$  & rel.(\%) $\uparrow$ &  abs.$\downarrow$ & rel.(\%)$\uparrow$  \\
\midrule
                    \texttt{ST1}& Context-T                           & \color{gray}{6.84}        &              & \color{gray}{4.95}         &              & \color{gray}{11.72}        &   \\
                    \texttt{ST2}& Cascaded \cite{cascaded1}           & 6.81                      &  (0.44)      & 5.07                       & (-2.42)      & 11.89                      & (-1.45)\\
                    \texttt{ST3}& \textbf{Joint optimization (ours)}  & \textbf{6.67}             &  (2.49)      & \textbf{4.86}              & (1.81)       & 11.79                      & (-0.60)\\
                    \texttt{ST4}& \textbf{S3 + sp-DD (ours)}          & 6.84                      &  (0.00)      & 4.95                       & (0.00)       & \textbf{11.62}             & (0.85)\\
                    \texttt{ST5}& \textbf{S3 + sp-ED (ours)}          & \textbf{\underline{6.66}} &  (2.63)      & \textbf{\underline{4.75}}  & (4.04)       & \textbf{\underline{11.10}} & (5.29)\\
\midrule
\midrule
                    \texttt{NST1}& Conformer-T                       & \color{gray}{5.78}        &              & \color{gray}{4.15}         &              & \color{gray}{9.94}         & \\
                    \texttt{NST2}& Cascaded \cite{cascaded1}         & 5.60                      &  (3.11)      & 4.04                       & (2.65)       & 9.68                       & (2.61)\\
                    \texttt{NST3}& \textbf{Joint optimization (ours)}& \textbf{5.48}             &  (5.19)      & \textbf{3.89}              & (6.27)       & \textbf{9.50}              & (4.42)\\
                    \texttt{NST4}& \textbf{NS3 + sp-DD (ours)}       & \textbf{5.41}             &  (6.40)      & \textbf{\underline{3.77}}  & (9.16)       & \textbf{9.38}              & (5.63)\\
                    \texttt{NST5}& \textbf{NS3 + sp-ED (ours)}       & \textbf{\underline{5.30}} &  (8.30)      & \textbf{3.87}              & (6.22)       & \textbf{\underline{8.98}}  & (9.65)\\
\bottomrule
\end{tabular}
}
\end{center}
\vspace*{-5mm}
\end{table}

\section{Experiments}
\noindent\textbf{Corpus.}
Our main results utilize a subset (subset A) of the Corpus of Spontaneous Japanese (CSJ) \cite{csj}, an academic lecture-based ASR task, consisting of 236 hours of speech. The evaluation set is divided into three tasks; eval 1, eval 2, and eval 3, containing 1.9 hours, 2.0 hours, and 1.3 hours of speech data, respectively.
\begin{table*}[t!]
\caption{Librispeech 100: WER(\%) and WERR(\%) and CSJ APS: CER(\%) and CERR(\%) with (i) conventional mean square error-based encoder-side distillation (mse-ED), (ii) proposed similarity-preserving-based decoder-side distillation (sp-DD), and (iii) proposed similarity-preserving-based encoder-side distillation (sp-ED). Streaming results are presented from \texttt{ST1-ST3} and non-streaming from \texttt{NST1-NST3}. Best WER/CER result is \textbf{bolded} and best overall results are further \underline{\textbf{underlined}}.}
\vspace*{-5mm}
\label{ablation2}
\begin{center}
\scalebox{0.83}{
\begin{tabular}{@{}llc@{\hspace{0.01cm}}cc@{\hspace{0.01cm}}cc@{\hspace{0.01cm}}cc@{\hspace{0.01cm}}c|c@{\hspace{0.01cm}}cc@{\hspace{0.01cm}}cc@{\hspace{0.01cm}}c}
\toprule
 &       & \multicolumn{8}{c|}{\textbf{LIBRISPEECH (100 H)}}       & \multicolumn{6}{c}{\textbf{CSJ (SUBSET A)}}  \\
 \cmidrule(lr){3-16}  
\textbf{Mode} & \textbf{Method}        & \multicolumn{2}{c}{\textbf{dev-clean}} & \multicolumn{2}{c}{\textbf{dev-other}} & \multicolumn{2}{c}{\textbf{test-clean}}              & \multicolumn{2}{c|}{\textbf{test-other}}         & \multicolumn{2}{c}{\textbf{eval1}}   & \multicolumn{2}{c}{\textbf{eval2}}       & \multicolumn{2}{c}{\textbf{eval3}}\\
 \cmidrule(lr){3-16}
&        &   abs.$\downarrow$ & rel.(\%)$\uparrow$ &   abs.$\downarrow$ & rel.(\%)$\uparrow$ &   abs.$\downarrow$ & rel.(\%)$\uparrow$ & abs.$\downarrow$  & rel.(\%) $\uparrow$ & abs.$\downarrow$  & rel.(\%) $\uparrow$  & abs.$\downarrow$  & rel.(\%) $\uparrow$             & abs.$\downarrow$  & rel.(\%) $\uparrow$\\
\midrule 
\texttt{ST1}  & Joint optimization + mse-ED  & 8.55          &        & 22.53            &          & 8.97           &              &  23.40             &               &  6.91             &         &  4.98             &         &  11.88           & \\
\texttt{ST2}  & Joint optimization + sp-DD   & 8.14          & (4.80) & 21.64            &  (3.95)  & 8.73           &  (2.68)      &  22.82             &  (2.48)       &  6.84             & (1.01)  &  4.95             & (0.60)  &  11.62           & (2.19) \\
\texttt{ST3}  & Joint optimization + sp-ED   & \textbf{\underline{8.14}} & (4.80) & \textbf{\underline{21.52}}   &  (4.48)  & \textbf{\underline{8.71}}  &  (2.90)      &  \textbf{\underline{22.73}}    &  (2.86)       &  \textbf{\underline{6.66}}    & (3.61)  &  \textbf{\underline{4.75}}    & (4.62)  &  \textbf{\underline{11.10}}  & (6.57)  \\
\midrule
\midrule
\texttt{NST1} & Joint optimization + mse-ED   & 7.15          &          & 19.85          &         & 7.44           &              &  20.70              &               &  5.69              &         &  3.93            &          &  9.79           &\\
\texttt{NST2} & Joint optimization + sp-DD    & 6.57          &  (8.11)  & 19.04          &  (4.08) & \textbf{\underline{6.90}}  &  (7.25)      &  19.58              &  (5.41)       &  5.41              & (4.92)  &  \textbf{\underline{3.77}}   & (4.07)  &  9.38           & (4.19) \\
\texttt{NST3} & Joint optimization + sp-ED    & \textbf{\underline{6.41}} &  (10.35) & \textbf{\underline{18.68}} &  (5.89) & 6.91           &  (7.12)      &  \textbf{\underline{19.16}}     &  (7.44)       &  \textbf{\underline{5.30}}     & (6.85)  &  3.87            & (1.53)     &  \textbf{\underline{8.98}}  & (8.27)\\
\bottomrule
\end{tabular}
}
\end{center}
\vspace*{-6mm}
\end{table*}

\textbf{Implementation details.} We develop our baseline models and the proposed architecture using the ESPnet2 toolkit \cite{espnet}. We introduce extra encoder and decoder modules to develop this particular multi-decoder model using existing specialized architectures. In our experiments, the streaming encoder module has twelve 256-dimensional contextual block conformer layers with 1024 feed-forward dimensions and 4 attention heads with a dropout rate of 0.1 for each of them. For block-processing \cite{streaming-encoder1}, we use a block size of 40 and keep the look-ahead and hop size 16 for optimal streaming performance. Similarly, for the non-streaming encoder, we use twelve 256-dimensional conformer layers with 1024 feed-forward dimensions and 4 attention heads. The output of the online contextual block conformer is provided as an input to the non-streaming conformer block to obtain the joint architecture. We use one separate RNN-T decoder for the streaming module with a 256-dimensional embedding prediction network and a 320-dimensional joint network. However, for the non-streaming module, we use four decoders as studied in \cite{multitask3} i.e., CTC, RNN-T, attention and MLM-based decoders. The CTC decoder uses a single linear layer, non-streaming RNN-T decoder comprises of a 256-dimensional prediction network and a 320-dimensional joint network. In contrast, attention and MLM-based decoders use six 512-dimensional attention heads and 256-dimensional feed-forward units. We train the joint architecture for 50 epochs with a learning rate of 0.0015 and warmup steps of 1500. In addition, we use an online training weight of 1 to maximize the performance capacity of the streaming module in joint training. For the non-streaming module, we adopt the training weights for \begin{math}\lambda_{\text{ctc}}, \lambda_{\text{att}}, \lambda_{\text{rnnt}},\end{math} \begin{math} \lambda_{\text{mlm}}\end{math} as proposed in \cite{multitask3}, i.e., 0.15, 0.30, 0.10 and 0.45 respectively. As this work also proposes to use knowledge distillation to preserve the similarities between the intermediate streaming and non-streaming encoder and decoder representations and uses the distillation weight \begin{math} \lambda_{\text{dist}} \end{math} of 3000 which is the optimal weight adopted from \cite{kd1}.
\vspace*{-3mm}
\subsection{Main results}
We compare our proposed jointly optimized multi-decoder E2E-ASR single model with the baseline standalone streaming contextual block conformer transducer (Context-T) and non-streaming conformer transducer (Conformer-T). Furthermore, we also implemented the baseline cascaded architecture with a shared RNN-T decoder as proposed in \cite{cascaded1}. For a fair comparison, we keep the Context-T and Conformer-T encoder layers same for all the experiments, i.e., twelve streaming and non-streaming encoder layers for the baseline and proposed architectures. The experiment results on the CSJ dataset are shown in Table \ref{ablation1}, where Context-T (\texttt{ST1}) and Conformer-T (\texttt{NST1}) are the standalone streaming and non-streaming transducer baseline models. A baseline cascaded architecture with shared decoders (\texttt{ST2} \& \texttt{NST2}) \cite{cascaded1} is compared with the proposed jointly optimized architecture (\texttt{ST3} \& \texttt{NST3}), decoder-side distillation (\texttt{ST4} \& \texttt{NST4}) and encoder-side distillation (\texttt{ST5} \& \texttt{NST5}). The character error rates (CER) presented in Table \ref{ablation1} are decoded using a beam width of 10 for streaming and non-streaming RNN-T paths. The results show that the proposed jointly optimized E2E-ASR model improved the performance of the streaming module compared to the standalone streaming model with a relative CER (CERR) improvement from 2.6\%-5.3\% in multiple evaluation sets and also showed improved performance for the non-streaming transducer path with 8.3\%-9.7\% CERR.

\textbf{Impact of knowledge distillation.}
We examine the effect of knowledge distillation using separate decoders in this study. We observe promising CERR in almost all the evaluation sets. While we have presented these results primarily from the perspective of enhancing the performance of the streaming module, which serves as the student network, our findings also indicate that preserving similarities within the hidden activations and utilizing them as an auxiliary loss function in joint training can also bring a regularization effect in the non-streaming module. This effect is brought by the joint architecture of the model, which can propagate improvements across both modules and takes advantage of the complementary similarity information. This observation is notable when compared to the baseline architectures, as seen in Table \ref{ablation1}.

\begin{figure}[t!]
    \centering
    \hfill
    \begin{subfigure}{0.23\textwidth}
        \includegraphics[width=\textwidth]{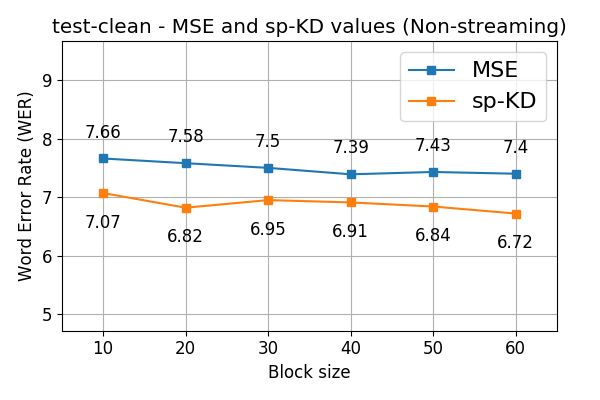}
        \caption{test-clean (non-streaming)}
        \label{fig:non_stream_test_clean}
    \end{subfigure}
    \hfill
    \begin{subfigure}{0.23\textwidth}
        \includegraphics[width=\textwidth]{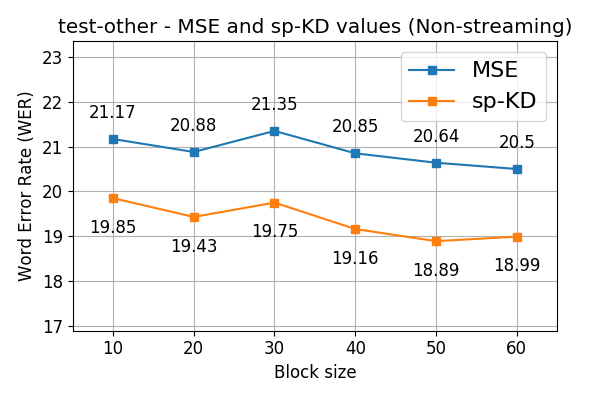}
        \caption{test-other (non-streaming)}
        \label{fig:non_stream_test_other}
    \end{subfigure}
    \hfill
    \begin{subfigure}{0.23\textwidth}
        \includegraphics[width=\textwidth]{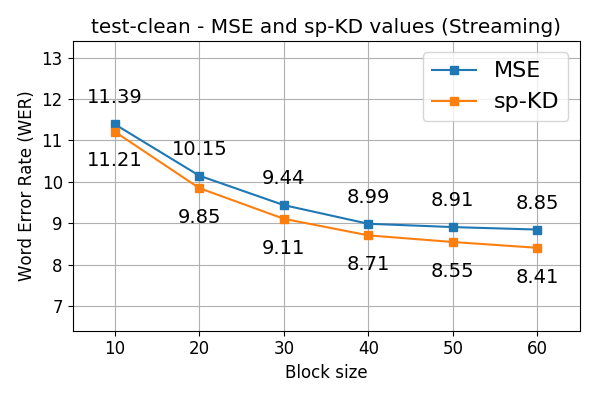}
        \caption{test-clean (streaming)}
        \label{fig:stream_test_clean}
    \end{subfigure}
    \hfill
    \begin{subfigure}{0.23\textwidth}
        \includegraphics[width=\textwidth]{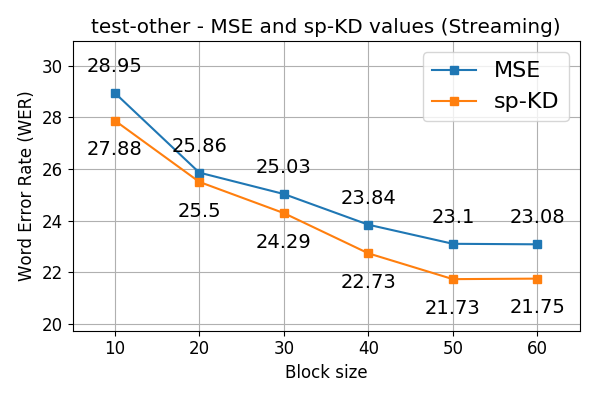}
        \caption{test-other (streaming)}
        \label{fig:stream_test_other}
    \end{subfigure}
\vspace*{-2mm}
    \caption{Comparative analysis using two knowledge distillation methods: mean square error-based encoder-side distillation (mse-ED) and similarity-preserving encoder-side distillation (sp-ED) (ours). Results are presented on two evaluation sets: test-clean, and test-other, for varying block sizes trained on Librispeech 100-hour dataset.}
    \label{fig:main_figure}
\vspace*{-6mm}
\end{figure}

\textbf{Comparative Analysis of Knowledge Distillation.}
In this study, we contrast the effectiveness of our proposed similarity preserving encoder-side (sp-ED) and decoder-side distillation (sp-DD) method with the conventional mean square error based encoder-side distillation (mse-ED) approach. This comparison takes place within our proposed jointly optimized multi-decoder architecture, across datasets such as Librispeech (100 hours) and CSJ (subset A). Table \ref{ablation2} and Figure \ref{fig:main_figure} explain these comparative results. In Table \ref{ablation2}, we focus on the performance of mse-ED, sp-DD and sp-ED methods on the Librispeech 100-hour and CSJ subset A datasets, using a block size of 40 in contextual layers. Figure 2 provides a more detailed analysis, highlighting the effectiveness of the sp-ED method compared to mse-ED in two modes: streaming and non-streaming and considers various block sizes, ranging from 10 to 60. We found that the sp-ED method consistently outperformed the mse-ED method, registering lower values across all datasets and block sizes, indicative of superior performance. For instance, in the dev-clean dataset, sp-ED reduces the WER to 6.41 at block-40 compared with the mse-ED method's lowest of 7.15. This trend held true even in streaming mode, where sp-ED's performance remained superior. Beyond the direct comparison of the sp-ED and mse-ED methods, it's essential to emphasize the significant role played by the joint optimization in the cascaded architecture in these results. The design, particularly the use of knowledge distillation in encoder layers, is integral in advancing improvements to the offline conformer module. This process subsequently boosts the performance of both streaming and non-streaming modes.

\textbf{Impact on Emission Delay.}
In Table \ref{tab:emissiondelay}, we compare the emission delay (ED) using 1 thread of CPU (AMD EPYC 7742) between the proposed model with knowledge distillation and the separate Context-T model. The proposed method maintains similar ED across block size of 20, with a slight increase at a block size of 40 and 60.
\begin{table}[h]
\vspace*{-5mm}
\caption{Comparison of emission delay for different block sizes.}
\vspace*{-5mm}
\label{tab:emissiondelay}
\begin{center}
\scalebox{0.8} {
\begin{tabular}{@{}cccc}
\toprule
Block size  & Block length (ms) & \multicolumn{1}{c}{Separate ED (ms)} & \multicolumn{1}{c}{Proposed ED (ms)}\\
\hline
20 & 800 & 125 & 125\\
40 & 1600 & 175 & 182\\
60 & 3200 & 200 & 231\\
\bottomrule
\end{tabular}
}
\end{center}
\vspace*{-11mm}
\end{table}
\vspace*{-2mm}
\vspace*{-0mm}
\section{Conclusion}
\vspace*{-3mm}
In this work, we have presented joint optimization in the multi-decoder-based network – an optimized architecture to combine streaming and non-streaming modules. This work leverages multi-task learning and uses separate decoders and intermediate encoder representations to improve the performance of both the modules while providing more flexibility in the E2E-ASR system. Compared to the specialized and conventional cascaded architectures \cite{cascaded1}, this work shows promising improvement in the CERR for CSJ corpus, especially with the introduction of intermediate similarity preserving knowledge distillation loss. 
\bibliographystyle{IEEEbib}
\bibliography{refs}

\end{document}